\documentstyle[prb,aps]{revtex}
\begin{document}
\draft\tighten
\title{Zero field muon spin lattice relaxation rate in a Heisenberg ferromagnet
at low temperature}
\author{P. Dalmas de R\'eotier and A. Yaouanc}
\address{Commissariat \`a l'Energie Atomique, \\ D\'epartement de Recherche
Fondamentale sur la Mati\`ere Condens\'ee, \\ F-38054 Grenoble cedex 9,
France}
\date{\today} \maketitle
\begin{abstract}
We provide a theoretical framework to compute the zero field muon spin
relaxation rate of a Heisenberg
ferromagnet at low temperature. We use the linear spin wave approximation. The
rate, which is a measure of the spin lattice relaxation induced by the
magnetic fluctuations along the easy axis,
allows to estimate the magnon stiffness constant.\par
\end{abstract}

%
\section{INTRODUCTION} \label{intro}
Recently, the measured temperature behavior of the zero field
muon spin relaxation ($\mu$SR) rate
of some Heisenberg ferromagnets has been analysed
quantitatively in the paramagnetic \cite{Yaouancepl,Dalmasprl,Dalmasprb} and
ferromagnetic \cite{Yaouancfutur} critical regimes. This analysis has given
information on the anisotropy of the spin dynamics in the reciprocal and
direct spaces. It has
shown that the anisotropy in the reciprocal space, which is induced by the
dipolar interaction, has a strong influence on the dynamics near the zone
center
of the Brillouin zone. Note that this interaction always exists in real
ferromagnets. The experimental data in the isotropic dipolar Heisenberg regime
are well explained by the mode coupling approximation of
Frey and Schwabl.\cite{Frey}
On the other hand, up to now, the low temperature behavior of
the relaxation rate has only been considered qualitatively\cite{Yaouancuk} or
in restricted physical cases.\cite{Loveseymuon}
The main purpose of this work is to provide a
theoretical framework to analyse $\mu$SR relaxation data recorded on
Heisenberg ferromagnets
at low temperature.\par
The organization of this paper is as follows. In Sec. \ref{muonlattice},
we summarize the relation between the measured depolarization function and the
spin-spin correlation function of the magnet.
In Sec. \ref{simple} we compute the muon spin lattice relaxation rate in the
linear
spin wave approximation for a Heisenberg hamiltonian with an energy gap.
In Sec. \ref{complet} we apply the result of the previous section to describe
the spin lattice relaxation of a dipolar Heisenberg magnet in the spin wave
temperature region. In this last section we present the conclusion of our
work.\par

\section{MUON SPIN RELAXATION AND SPIN-SPIN CORRELATION
FUNCTIONS}\label{muonlattice}

We take the $Z$-axis parallel to the incoming muon beam polarization. A
zero field muon spin relaxation measurement consists of measuring
the time
depolarization function $P_Z(t)$. \cite{Chappert,Schenck}
For simplicity we take the easy magnetic axis $z$ parallel to $Z$.
We write \cite{McMullen,Dalmasuk}
\begin{mathletters}
\label{depol}
\begin{eqnarray}
P_Z(t) & = &\exp \left[-\psi_z(t)  \right], \label{eq1:1}
\end{eqnarray}
with
\begin{eqnarray} \psi_z(t) & = &{\gamma^ 2_\mu} \int^{ t}_{0}
d\tau \ (t-\tau)\left[\Phi_{ xx}(\tau)+\Phi_{ yy}(\tau) \right]. \label{eq1:2}
\end{eqnarray}
\end{mathletters}
$\gamma_\mu$ is the muon gyromagnetic ratio : $ \gamma_ \mu$ = 851.6
Mrad s$^{ {\rm -1}}$T$^{-1} $.
$\psi_z(t)$ does not depend on the muon pulsation frequency
$\omega_\mu$ because the associated energy $\hbar\omega_\mu$
is negligible. \cite{Yaouancuk} In Eq. (\ref{eq1:2})
$\Phi_{ \alpha \alpha} (\tau ) $ is the
symmetrized time correlation function of the $\alpha$ fluctuating component of
the local magnetic field at the muon site and $\{x,y,z\}$ is an orthogonal
frame.\par
Experimentally it is usually observed that $P_Z(t)$ is an exponential function.
Eq. (\ref{depol}) predicts
such a result if $\tau$ is very small relative to $t$, {\it i.e.} if the
characteristic decay time of $\Phi_{ \alpha \alpha} $ is much smaller than
$t$ which
is $\sim$ 2 $\mu$s.
Taking this hypothesis which is justified and the fact that the field
correlation functions are
even functions of $\tau$, we
derive that $P_Z(t)$ is an exponential function characterized by a
damping rate $\lambda_z$ which can be written in terms of the time Fourier
transform of field correlation functions at $\omega$ = 0
\begin{eqnarray}
\lambda_ z & = & \pi \gamma^ 2_\mu \left[\Phi_{ xx}(\omega
=0)+\Phi_{ yy}(\omega =0 ) \right].
\label{time_fourier}
\end{eqnarray}
It is possible to express the
magnetic field at the muon site as a function of, on one hand, a tensor
which describes the coupling
between the localized spins of the metal and the muon spin and on the other
hand of the localized spins components themselves. Because
$ \Phi_{ \alpha \alpha} (\tau )$ is quadratic in fields, $ \lambda_ z $
can be written as a
linear combination of spin-spin correlation functions of the magnet. In general
the calculation of these correlation functions is performed in
$\bf q$ space, {\it i.e.} in the first Brillouin zone. Therefore
the spatial Fourier transform of these functions has to be introduced. This is
done in detail in Ref.~\onlinecite{Yaouancepl}. Using the result given at
Eq. (\ref{time_fourier}) we find
\begin{eqnarray}
\lambda_z & = & {\pi{\cal D} \over V} \int^{ }_{ }{ d^3 {\bf q}
\over( 2\pi)^ 3} \sum^{ }_{ \beta \gamma} \left[G^{x\beta} ( {\bf q})
G^{x\gamma} (- {\bf q} ) \right. \cr  & & \left. +G^{y\beta} ( {\bf q})
G^{y\gamma} (- {\bf q} ) \right] \tilde  \Lambda^{ \beta \gamma}
( {\bf q} ).
\label{general}
\end{eqnarray}
The integral extends over the first Brillouin zone. This equation shows that
$\lambda_z$ depends on the coupling between the muon spin and the spins of
the magnet through the coupling tensor $G({\bf q})$ and the spin
correlation tensor of the magnet, $\Lambda ( {\bf q} ,\omega ) $,
taken at $\omega$ = 0, {\it i.e.} $ \tilde  \Lambda ( {\bf q} ) $ $\equiv$
$ \Lambda ({\bf q} ,\omega =0) $. The information on the muon
localization site is contained in $G({\bf q})$.
We have defined $ {\cal D}= \left(\mu_ 0/4\pi
\right)^2\gamma^ 2_\mu \left(g_L\mu_ B \right)^2\  $where $ \mu_ 0\  $is the
permeability of free space, $g_L$ the Land\'e factor
and $\mu_ B$ the Bohr magneton. $V$ is the volume of the sample.
The formula given at Eq. (\ref{general}) allows to deal with Bravais and non
Bravais lattices, for instance the hexagonal close packed lattice. In the
latter case $G({\bf q})$ = $1/n_d \sum_{d} G_d ({\bf
q})$ where the index $d$ runs over the $n_d$ non equivalent sites of the
magnet and
$G_d({\bf q})$ = $\sum_{i} \exp [i{\bf q}\cdot({\bf i} + {\bf d}) ] G_{{\bf
r}_{i+d}}$. $i$ runs over each cell of the crystal lattice and $G_{{\bf
r}_{i+d}}$ is a dimensionless tensor which accounts for the classical
dipolar and Fermi contact couplings between the muon spin and
the magnet spins located at distance vector ${\bf r}_{i+d}$ from the muon.
$\tilde \Lambda ({\bf q})$ writes $\sum _{d,d^\prime} \tilde
\Lambda_{d d^\prime} ({\bf q})$ where $\tilde \Lambda_{d d^\prime} ({\bf q})$
is the correlation tensor between spins belonging to cell sites $d$ and
$d^\prime$. Note that the case of Bravais lattice is included in the definition
we have just given : $d$ takes then only one value ($n_d$ = 1) and the sums
reduce to a single element. \par
The tensor $G({\bf q})$ is the sum of dipolar and hyperfine tensors
which we note $D({\bf q})$ and $H({\bf q})$ respectively. In
general terms, the sublattice $d$ contribution to $D({\bf q})$ writes
\cite{Yaouancepl}
\begin{mathletters}
\label{ewald}
\begin{eqnarray}
 D^{\alpha \beta}_d( {\bf q}) & = &
-4\pi \left[P^{\alpha \beta}_ L( {\bf q} )-C^{\alpha \beta}_d ( {\bf q} )
\right],
\label{ewald1:1}
\end{eqnarray}
where $P^{\alpha \beta}_ L( {\bf q} )$ is the longitudinal projection
operator and C$^{\alpha \beta}_d ( {\bf q} )$ a symmetric tensor :
\begin{eqnarray}
C^{\alpha \beta}_d ( {\bf q} ) & = &
{ q_\alpha q_\beta \over q^2} \left[1-\exp \left({-q^2
\over 4\varrho^ 2} \right) \right]
- {1 \over 4\varrho^ 2} \sum^{}_{ {\bf K} \not= {\bf 0}}
\left(K_\alpha +q_\alpha \right) \left(K_\beta
+q_\beta \right)\varphi_ 0 \left({( {\bf q} + {\bf K})^2 \over 4\varrho^ 2}
\right)\exp \left(-i {\bf K} \cdot {\bf r}_{0+d} \right)   \cr
& & +{v \varrho^ 3
\over 2(\pi)^{3/2}} \sum^{ }_ i \left[2\varrho^ 2r_{i+d,\alpha} r_{i+d,\beta}
\varphi_{ 3/2} \left(\varrho^ 2r^2_{i+d} \right)-\delta^{\alpha\beta}\varphi_{
1/2} \left(\varrho^ 2r^2_{i+d} \right) \right]\exp \left(i {\bf q} \cdot {\bf
r}_{i+d} \right).
\label{ewald1:2}
\end{eqnarray}
\end{mathletters}
The $D({\bf q})$ expression is derived using an Ewald transformation.
\cite{Born} The $\varphi_m(x)$ functions are defined in
Ref.~\onlinecite{Yaouancepl}. $\bf K$ is a vector of the reciprocal lattice
and ${\bf r}_{i+d}$ (respectively ${\bf r}_{0+d}$) is the vector that
links the muon site to the ion belonging to sublattice $d$ and located in
lattice cell $i$ (respectively origin) of the crystal lattice.
Expression
(\ref{ewald1:2}) gives the same result for all values of the Ewald parameter
$\varrho$, but for numerical applications a value of  $\varrho$ is chosen which
ensures that both series of Eq. (\ref{ewald1:2}) converge rapidly.
The $C_d(\bf q)$ tensor, the trace of which is 1, reveals the symmetry of
the point group at the muon site.
Whereas the elements of the $C_d({\bf q})$ tensor are analytical functions
of $\bf q$ for all $\bf q$ values, the elements of $P_L({\bf q})$ are only
piecewise continous at ${\bf q}=0$. \par
The hyperfine interaction is short range and usually isotropic. In the
lowest order in $\bf q$ we have
\begin{eqnarray}
H^{\alpha\beta}({\bf q}=0) & \equiv & 1/n_d \sum_d H^{\alpha\beta}_d({\bf q}=0)
= r_\mu H \delta^{\alpha\beta},
\label{hyperfine}
\end{eqnarray}
where $r_\mu$ is the number of nearest neighbor magnetic ions to the muon site
and $H$ a constant which can be deduced from the muon spin rotation
frequency at low temperature.
Equation (\ref{hyperfine}) is derived under the hypothesis that the muon site
is a center of symmetry. \par
In order to proceed further, in this paper we will only consider the behavior
of $ G( {\bf q})$
near the zone center. From the previous results, using
$C^{\alpha\beta}({\bf q})$ $\equiv$ $1/n_d \sum_d C^{\alpha\beta}_d({\bf q})$,
we derive in the limit ${\bf q} \rightarrow 0$
\begin{eqnarray}
G^{\alpha \beta}( {\bf q}) & = & -4\pi \left[
\vphantom{r_\mu H\delta^{\alpha\beta}\over{4\pi}}
P^{\alpha \beta}_ L( {\bf q} )-C^{\alpha \beta} (0)
- {r_\mu H\delta^{\alpha\beta}\over 4\pi}\right].
\label{coupling}
\end{eqnarray}

\section{MUON SPIN RELAXATION RATE IN AN HEISENBERG FERROMAGNET WITH A
SMALL ENERGY GAP}\label{simple}

For the computation of the damping rate we need an expression
for the correlation functions of the magnet.
We first have to specify its hamiltonian. As we have done before
\cite{Yaouancuk} we suppose that the magnet is described by a Heisenberg
interaction with a small energy gap.
Because the minimum magnon energy is much larger than $\hbar\omega_\mu$, the
energy
conservation principle tells us that only the parallel (to
the easy axis $z$) fluctuations contributes to the depolarization,
\cite{Yaouancuk} {\it i.e.} the measurements probe only
the correlation function $\tilde  \Lambda^{ zz}({\bf q})$.
Then a close look at Eq. (\ref{general}) indicates that the
diagonal terms of $G({\bf q})$ do not contribute to the depolarization.
This means
that an isotropic hyperfine interaction does not influence the measured damping
rate. For many possible muon localization sites in crystals we have
$ C^{zx} ({\bf q}=0) $ = $ C^{zy} ({\bf q}=0) $ = 0. Using this hypothesis
we deduce the simple
result
$G^{z \beta}( {\bf q} \rightarrow 0)$ =
$- 4 \pi P^{z \beta}_ L( {\bf q} )$ for $\beta$ = $x$, $y$.
This leads to the important fact that,
within our hypothesis, the muon spin relaxation rate is independent of the
muon localization site. The previous analysis
leads to the following simple result :
\begin{eqnarray}
\lambda_ z & = & {2{\cal D} \over V} \int^{ }_{ } d^3
{\bf q}{q^2_z
\over q^2} \left(1-{q^2_z \over q^2} \right)\tilde  \Lambda^{ zz}(\bf q).
\label{lambdaint}
\end{eqnarray}
A generalization of this result when $C^{zx}({\bf q}=0)$ and
$C^{zy}({\bf q}=0)$ are not zero is not
difficult : it results in a modified prefactor that depends on
$C^{zx}({\bf q}=0)$ and $C^{zy}({\bf q}=0)$.
We compute $ \tilde  \Lambda^{zz} ({\bf q}) $ in the linear
spin wave approximation. \cite{Holstein}
In this approximation the
fluctuating part of the $z$ component of the total angular momentum of the
magnetic ion,
$\delta{\bf J}_{\bf q}^z$, is expressed in terms of a sum of products of
boson operators $a^+_{\bf k}$ and $a^-_{\bf k}$:
\begin{eqnarray}
\delta{\bf J}_{\bf q}^z & = & -{1\over N}
\sum^{ }_{\bf k,k_1}\delta({\bf {-q+k-k_1,K}})a^+_{\bf k}a^-_{\bf k_1},
\label{jzmagnon}
\end{eqnarray}
where ${\bf K}$ is a vector of the reciprocal lattice. Each of the
$\bf k$ and $\bf k_1$ sums
extends over $N$ vectors of the first Brillouin zone.
Because $\delta{\bf J}_{\bf q}^z$ is given in terms of products of two
operators,
the computation of $\tilde  \Lambda^{ zz}(q)$ requires the evaluation of
four operator products. We decouple these products. We set :
\begin{eqnarray}
<a^+_{\bf k}a^-_{\bf k_1}a^+_{\bf q}a^-_{\bf q_1}> & = &
N^2\delta_{{\bf k_1}{\bf q}} \delta_{{\bf k}{\bf q_1}}n_{\bf k}(n_{\bf q}+1),
\label{fock}
\end{eqnarray}
where $n_{\bf q}$ is the standard Bose occupation factor. With these results
it is easy to derive an expression for the parallel spin-spin
correlation function. We find :
\begin{eqnarray}
\tilde  \Lambda^{zz}(\bf q) & = &\sum^{ }_ {\bf p}
{{\exp({\hbar\omega_{\bf p}}/{k_BT})}
\over{[\exp({\hbar\omega_{\bf p}}/{k_BT})-1]^2}}
{\delta\left(\omega_{\bf {p+q}}- \omega_{\bf p}\right)}.
\label{partiel}
\end{eqnarray}
In order to proceed further we need an expression for the magnon dispersion
relation $\hbar\omega_{\bf q}$. The simplest choice is to write
\begin{eqnarray}
\hbar\omega_{\bf q} & = & D_m q^2+ \Delta,
\label{magnon}
\end{eqnarray}
where $D_m$ is the magnon stiffness
constant and $\Delta$ is the energy gap of the magnon dispersion relation.
We have supposed that the dispersion relation is isotropic at small $\bf q$.
The sum in Eq. (\ref{partiel})
can be converted to an integral and can
be computed analytically using Eq. (\ref{magnon}). We obtain :
\begin{eqnarray}
\tilde  \Lambda^{ zz}(q) & = &{V\hbar \over 16\pi^ 2} {k_BT \over
D_m^2 q} \left[
{1 \over \exp \left[ \left(D_mq^2 /4+\Delta \right)
/k_BT \right]-1}
\right. \cr
& & \left.
- {1 \over \exp \left[ \left(D_m q^2_{BZ} +\Delta \right)
/ k_BT \right]-1 }\right],
\label{correlation}
\end{eqnarray}
where $ k_B $ is the Boltzmann constant and $q_{BZ}$ the radius of the
Brillouin zone. We note that usually at low temperature we have
$ \left(D_m q^2_{BZ}+\Delta \right)\gg k_B T $. This is the case, for instance,
for GdNi$_5$ for which $D_m$ = 3.2 meV.\AA$^2$ and $q_{BZ}$ =
0.9 \AA$^{-1}$ (see Ref.~\onlinecite{Yaouancfutur}). Then the expression of
the correlation function greatly simplifies :
\begin{eqnarray}
\tilde  \Lambda^{ zz}(q) & = &{V\hbar \over 16\pi^ 2} {k_BT \over
D_m^2q} \cr
& & \times {1 \over \exp \left[ \left(D_mq^2 \left/4+\Delta \right. \right)
\left/k_BT \right. \right]-1}.
\label{correlationsimple}
\end{eqnarray}
Notice that, because of the simple relation we take for the magnon dispersion
relation, the correlation function is a function of  the magnitude of $\bf q$
and not of its orientation. \par
So far in this section we have not mentioned if we are dealing with Bravais or
non Bravais lattices. In a non
Bravais lattice where $\tilde \Lambda^{zz} ({\bf q})$ writes
$\sum _{d,d^\prime} \tilde \Lambda_{d d^\prime}^{zz} ({\bf q})$,
the existence of more than one atom per unit cell gives rise
to acoustic and optical branches in the dispersion relation. Because of the
relatively huge gap present in the optical branch(es), the spin correlation
functions $\tilde \Lambda^{zz}_{d d^\prime} ({\bf q})$ with $d$ $\not=$
$d^\prime$ are not relevant as seen at Eqs. (\ref{partiel}) or
(\ref{correlation}).\par
The expression of the muon spin relaxation rate for
the model considered in this section (specified by the dispersion relation
given at Eq. (\ref{magnon})) is obtained using Eqs. (\ref{lambdaint})
and (\ref{correlationsimple}). We
derive
\begin{mathletters}
\label{lambdacomplique}
\begin{eqnarray} \lambda_ z & = &{{\cal C}g^2_L T^{2}\over D_m^3}P(q_{BZ}),
\label{lambdacomplique:1} \end{eqnarray}
with
\begin{eqnarray} P(q) & = &
\ln \left[{1-\exp \left[-(D_m q^2/4+\Delta)/k_BT \right] \over
1- \exp \left(-\Delta/k_BT \right)} \right], \label{function} \end{eqnarray}
\end{mathletters}
where we have defined ${\cal C}$ =
$(2/15\pi) \left(\mu_ 0/4\pi \right)^2\gamma^ 2_\mu \mu^
2_{_B}\hbar k^2_B $ =
129.39 (meV)$^3$.\AA$^6$.s$^{-1}$.K$^{-2}$.
In the limiting case, $(D_m q^2_{BZ}/4 +\Delta)$ $>$ $D_m q^2_{BZ}/4$
$\gg$ $k_BT$, we have
\begin{eqnarray} \lambda_ z & = &{{\cal C}g^2_L T^{2}\over D_m^3}
\ln \left[{\exp \left(\Delta/k_BT \right) \over \exp \left(\Delta/k_BT
\right)-1} \right]. \label{lambda} \end{eqnarray}
A numerical study of the $P(q)$ function shows
that the expression for $\lambda_z$ given
at Eq. (\ref{lambda}) is a
reasonable approximation of the $\lambda_z$ expression given at Eq.
(\ref{lambdacomplique}).\par

In practice we have
$\Delta$ $\ll$ $k_BT$. In this high temperature regime we obtain the simple
result :
\begin{eqnarray} \lambda_ z & = &{{\cal C}g^2_L T^{2}\over D_m^3}
\ln \left({k_BT/\Delta} \right). \label{lambdauk} \end{eqnarray}
The $T^2 \ln(k_BT/\Delta)$ dependence of $\lambda_z$ has been predicted in
the past for an anisotropic contact interaction which can occur in nuclear
magnetic resonance. \cite{Mitchell,Beeman} Some years ago we have derived
this temperature dependence for muon spin relaxation rate using the
model considered in the present work.\cite{Yaouancuk} But the derivation
was not sound mathematically : the constant
$G$ (introduced at Eq. (15) of Ref.~(\onlinecite{Yaouancuk}) was
undetermined.\par
The physical origin of the $T^2$ factor in Eq. (\ref{lambdauk}) is clear~:
each of the two magnons contributing to the muon
spin depolarization process (Raman process) accounts for a factor $T$.
We note that $\lambda_z$ given by Eq. (\ref{lambdauk}) diverges if $\Delta$
= 0, {\it i.e.} for a pure Heisenberg magnet.
This fact is not disturbing because a real magnet has always a small energy
gap.\cite{Kittel}\par
Although the result of Eq. (\ref{lambdauk}) has not been published before, it
has already been used
to extract the stiffness constant of the dipolar ferromagnet GdNi$_5$ from
its zero muon spin relaxation rate data. \cite{Yaouancicm} \par

$\lambda_z$ has been computed by other authors \cite{Loveseymuon} for EuO. In
the present work, we have presented a general framework to compute the $\mu$SR
relaxation rate for a ferromagnet at low temperature. Then for a given
magnon dispersion relation the rate has
been computed explicitely. In the proposed framework
the coupling between the muon spin and the electronic spins is described by the
tensor $G(\bf q)$. For a given muon localization site in a crystal, its
elements are easily computed. One does not have to consider separately the
relaxation due to the hyperfine and dipolar interaction.\par

\section{MUON SPIN RELAXATION IN A DIPOLAR HEISENBERG FERROMAGNET AT LOW
TEMPERATURE}\label{complet}
In the previous section we have computed $\lambda_z$ for a simple
Heisenberg hamiltonian with an energy gap. In a dipolar magnet the
gap is induced by the dipolar interaction between the ions.
Recently the neutron scattering function has been computed for a dipolar
Heisenberg magnet in the linear spin wave approximation. \cite{Loveseyth}
{}From this work an expression for the $zz$ correlation
function can be deduced. Relative to our expression
the differences are that a relatively complicated weighting factor exists
besides the thermal factor in Eq. (\ref{partiel}) and the magnon dispersion
relation for a dipolar Heisenberg magnet is used. However, qualitatively
this factor does not influence the correlation function $\tilde
\Lambda^{zz}({\bf q})$. The magnon dispersion relation writes \cite{Keffer}
\begin{eqnarray}
\omega^2_{\bf q} & = & \varepsilon_{\bf q}(\varepsilon_{\bf q} +
\epsilon\sin^2 \theta_{\bf q}),
\label{dispersion}
\end{eqnarray}
where $\theta_{\bf q}$ is the polar angle of wave vector ${\bf q}$ relative to
the easy axis,
$\varepsilon_{\bf q}$ the dispersion relation without the dipolar interaction
and $\epsilon$ = ${(\mu_0g_L \mu_B M_0)}/\hbar$ a characteristic dipolar
energy ($M_0$ is the saturation magnetization). Therefore a natural choice
is to identify the gap $\Delta$ with $\hbar \epsilon$. In this case
the effect of $\sin^2 \theta_{\bf q}$ is neglected.
An other possibility is to take
for the gap value the bulk magnetic anisotropy energy :
$\Delta$ = $g_L \mu_B B_a$ where $B_a$ is the magnetic anisotropy field. These
two choices represent limiting cases. We note that for GdNi$_5$ we have
$\hbar \epsilon /k_B$ $\sim$ 1.2 K and $(g_L \mu_B B_a)/k_B$
$\sim$ 0.3 K, {\it i.e.}
there is a factor 4 difference. But because the purpose of the measurement of
the relaxation rate is to measure the spin wave stiffness constant which is
proportionnal to $[\ln (k_BT/ \Delta)]^{1/3}$, this uncertainty by a factor of
4 will only introduce an error of $\sim$ 20 $\%$ on the extracted $D_m$
value from the relaxation data (in the case of GdNi$_5$ the $D_m$
value is deduced from the relaxation data recorded at about 10 K - 15 K).\par
In this work we have given a general framework to analyse positive muon
longitudinal relaxation data at low temperature for a Heisenberg magnet.
Using the simplest possible approximation, we have shown that the measurements
allow to estimate the magnon stiffness constant. The framework which is given
here should allow to deal with other problems in an easy way. We
think for instance of antiferromagnets.\par

\acknowledgements{We thank A.V. Lazuta, S.V. Maleyev and A.G. Yashenkin
for useful discussions and P.C.M. Gubbens for his constant interest.}

\end{document}